\documentclass[preprint]{aastex}
\usepackage[]{natbib}
\usepackage[]{verbatim}
\usepackage{algorithm}
\usepackage{algorithmic}
\usepackage{amsmath}
\usepackage{graphicx}

\usepackage{etex}
\reserveinserts{18}
\usepackage{morefloats}
\usepackage{lscape}
\usepackage{color}
\usepackage{ulem}
\usepackage{amssymb}
\usepackage[caption=false]{subfig}    

\newcommand{\ergs}{${\rm ergs~s^{-1}}$}

\newcommand{\oiii}{[O\,{\sc iii}]}

\newcommand{\Msigma}{\ifmmode M_{\rm BH} - \sigma \else $M_{\rm BH} - \sigma$\fi}

\newcommand{\Halpha}{\ifmmode {\rm H}\alpha \else H$\alpha$\fi}

\newcommand{\Hbeta}{\ifmmode {\rm H}\beta \else H$\beta$\fi}
\newcommand{\herschel}{\textit{Herschel}}
\newcommand{\spitzer}{\textit{Spitzer}}
\newcommand{\galex}{\textit{GALEX}}

\shorttitle{star formation in AGN host galaxies}
\shortauthors{Xu et al.}
\begin{document}

\title{The Relation between Luminous AGNs and Star Formation in Their Host Galaxies}

\author{Xu, Lei\altaffilmark{1}, Rieke, G. H.\altaffilmark{1}, Egami, E.\altaffilmark{1},  Haines, C. P.\altaffilmark{1,3}, Pereira, M. J.\altaffilmark{1}, \& Smith, G. P.\altaffilmark{2}}

\altaffiltext{1}{ Steward Observatory, 933 N. Cherry Ave, University of Arizona, Tucson, AZ 85721, USA}
\altaffiltext{2}{School of Physics and Astronomy, University of Birmingham, Edgbaston, Birmingham B15 2TT, UK }
\altaffiltext{3}{Departamento de Astronomìa, Universidad de Chile, Casilla 36-D, Correo Central, Santiago, Chile}


\begin{abstract}
We study the relation of active galactic nuclei (AGNs) to star formation in their host galaxies. 
Our sample includes 205 Type-1 and 85 Type-2 AGNs, 162 detected with {\it Herschel}, from fields surrounding 30 galaxy clusters 
 in the {\bf Lo}cal {\bf C}l{\bf u}ster {\bf S}ubstructure {\bf S}urvey (LoCuSS).  
The sample is identified by optical line widths and ratios after selection to be brighter than 1 mJy at 
24$\mu$m. We show that Type-2 AGN [OIII]$\lambda$5007 line fluxes  
at high z can be contaminated by their host galaxies with typical spectrograph entrance apertures 
(but our sample is not compromised in this way). We use spectral 
energy distribution (SED) templates to  decompose the galaxy SEDs and estimate star formation rates, 
AGN luminosities, and host galaxy stellar masses (described in an 
accompanying paper). The AGNs arise from massive black holes ($\sim 3 \times 10^8 M_\odot$) accreting 
at $\sim$ 10\% of the Eddington rate and residing in
galaxies with stellar mass $> 3\times 10^{10}  \rm~ M_\odot $; those
detected with \herschel\/ have IR luminosity from star formation in the range of 
 $L_{\rm SF,IR} \sim 10^{10}-10^{12} \rm~ L_\odot$. 
We find that: 1.) the specific star formation rates in the host galaxies are 
generally consistent  with those of normal star-forming (main sequence) galaxies; 
2.) there is a strong correlation between the luminosities from star formation and
   the AGN; and 3.) however, the correlation may not result from a causal connection, but could arise because
   the black hole mass (and hence AGN Eddington luminosity)  and star formation
   are both correlated with the galaxy mass.

\end{abstract}


\section{Introduction}
\label{section:intro}

A key question in galaxy evolution is how the link is established and maintained between 
the mass of stars in a galaxy bulge and the 
mass of the super massive black hole (SMBH) in its center (e.g., Magorrian et al. 1998; Tremaine et al. 2002; Kormendy \& Ho 2013). 
To probe this issue, 
we need to understand in detail the connection between 
accretion and star forming activity over cosmic time in AGNs and their hosts \citep{heckman14}.

Because of the extreme luminosities of Type-1 AGN, they account for much of the accretion by SMBHs 
and therefore are critical to understanding this link. 
Their study has produced an overall understanding of the growth of mass in
nuclear black holes (e.g., \citet{marconi04, heckman14}).
The picture of star-forming galaxy evolution and
of the growth of stellar mass at $z < 2$ is also 
 becoming more clear. Star formation rates
(SFRs) in normal star-forming galaxies are proportional to  
the galaxy stellar mass, with specific  SFRs 
(SSFR, i.e., the star formation rate per unit stellar mass)
increasing with redshift (e.g., Noeske et al.
2007; Elbaz et al. 2007; Daddi et al. 2007). 
Typical vigorously star-forming galaxies lie on a  ``main sequence", where
star formation is likely induced by 
internal processes such as turbulence and disk instabilities. However, a significant fraction of
galaxies with stellar masses $\ge 3 \times 10^{10}$ M$_\odot$  
 have SSFRs well below the main sequence, which therefore needs to be
interpreted as an upper limiting case for the majority of star-forming galaxies. 
In addition, starbursting galaxies are a minority undergoing an episode of 
elevated star formation that places them well above the main sequence. 

Despite these advances, studying the co-evolution of stellar and SMBH mass growth is challenging. 
Since the host galaxies are often hidden in the UV, optical, and NIR  
by the strong glare of the AGN, using these bands to
compare the AGN host galaxy pattern of star formation with that in normal galaxies is very difficult.
The mid-infrared aromatic (hereafter PAH) feature luminosity is found to be correlated to 
other possible SFR indicators in the quasar spectra, such as far infrared (FIR) luminosity. 
The  PAH features  in quasar and radio galaxy spectra   
 at low-z \citep{schweitzer06,shi07, dicken12, shi14}, intermediate-z \citep{shi09},  and high-z 
\citep{maiolino,lutz08,shipley13} are direct evidence for  star formation. 
Further evidence of star formation in AGN hosts
comes from CO detection of quasars from low-z to high-z 
(e.g. Scoville et al. 2003, Solomon
\& Vanden Bout 2005, Wang
et al. 2010). The CO line luminosities are also correlated with the FIR 
luminosities (e.g. Riechers et al. 2011).
The correlations between FIR luminosity and these other star
formation tracers suggest that FIR measurements may offer a useful approach
for characterizing this process in the host galaxies. 

Prior to \herschel\/, the measurements of rest-frame FIR emission from AGNs were limited to 
a small population (e.g.,  Omont et al. 2001; Haas et al. 2003; Dicken et al. 2008).
With the advent of \herschel\/, it is possible 
to study  the FIR properties efficiently for much larger samples. 
The sensitivity and wavelength coverage of \herschel\/  (PACS: 100, 160 $\mu$m; SPIRE: 250, 350, 
and 500 $\mu$m)  can sample the
IR spectral energy distributions (SEDs) of AGNs up to $z \sim 3$.
A number of Type-1 AGN studies with  \herschel\/ have been reported. 
For example, \citet{hatziminaoglou10} present SPIRE data for 
a heterogeneously selected (from SDSS/MIR)
Type-1 AGN sample at $z\sim 0-3$. The SPIRE 3-band colors of their \herschel\/-detected
AGN are  indistinguishable from those of non-AGN star-forming galaxies. 
\citet{leipski13} publish \herschel\ data for eleven  1.2 mm-selected Type 1 AGNs at $z \sim 6$, 
and show that five emit strongly in the FIR. \citet{leipski14} extend this survey to a total of 69 systems, 
of which $\sim$ 30\% are detected in the FIR. 
These studies find that an AGN-powered torus is not enough 
to account for all the FIR emission of  \herschel-detected AGNs, 
and a star forming component is needed. 
\herschel\/ observations have also been reported
for X-ray selected, 
moderate luminosity AGNs at $z \sim 0.5-3$ 
(e.g., Shao et al. 2010; Mullaney et al. 2012). \citet{mullaney12}
find that the majority (79\%) of their \herschel-detected AGNs  reside in 
massive, normal star-forming galaxies, where the star formation
appears to be induced by  internal processes rather than by major mergers. 
However, no correlation is found
between the X-ray luminosity from AGN and the FIR luminosity powered by star formation. 
\citet{dicken12} used {\it Spitzer} data on nearby radio galaxies to show that there is 
no connection between star formation in the hosts and the nuclear activity. 
\citet{rosario13} studied a relatively luminous quasar 
sample and found the mean SFRs of the host galaxies were consistent 
with the galaxy `main sequence' and could be linked to the AGN properties through a common
dependence on stellar mass. However, this study relies on stacking to complement
the individual FIR detections, so their conclusions may not hold for a minority of galaxies with above-average
SFRs. In addition, the stacking results may be influenced by far-infrared-quiescent 
host galaxies. The presence of a significant quiesent host population seems likely, since 
local quasars can occupy quiescent hosts. For example, \citet{shi14} find that 16 of
84 PG quasars (or 19\%) have SFRs $\le$ 1 M$_\odot$/yr.

In most previous work,
the AGN sample was defined through optical spectroscopy (e.g., Hatziminaoglou et
al. 2010, Rosario et al. 2013), or X-ray emission (e.g., Rafferty et al. 2011; Mullaney et al. 2012).
Both of these metrics should emerge after obscuring circumnuclear material has cleared; that is,
they represent late-phase AGNs, if we assume an evolution from gas inflow to activating an AGN
to gas outflow (e.g., Hopkins 2012). Some previous studies are based on incomplete wavelength
coverage with \herschel\ (e.g. only SPIRE, Hatziminaoglou et al. 2010; or only PACS, Rosario
et al. 2013), thus limiting the accuracy with which the far infrared SEDs can be determined.
In addition, virtually all of these works center on moderate-luminosity AGN (e.g., Shao et al.
2010; Mullaney et al. 2012; Rosario et al. 2013), which may be substantially below the SMBH
Eddington luminosities and hence require modeling to relate the average (or peak) nuclear activity
to host galaxy properties (e.g., Chen et al. 2013; Hickox et al. 2013). Finally, in many studies the
sample of galaxies detected in at least two \herschel\ bands (supporting accurate SED fitting) is
relatively small (e.g., 49 in Hatziminaoglou et al. 2010, 38 in Rosario et al. 2013). Consequently,
the conclusions based on these previous works are sometimes inconclusive or contradictory.

We will augment and improve on these studies in a number of ways. 
The 205 Type-1 AGNs in our sample are uniformly selected from a 5.2 $\rm {deg}^2$ survey area, 
on the basis of 24 $\mu$m flux density above 1 mJy.  
We complement these objects 
with 85 Type-2 AGN, selected in the same way from part of the same area. We draw 15
galaxies from this latter sample that are directly comparable with the Type-1 AGNs in terms 
of black hole masses and accretion rates, and which fall within the same redshift range.
A multi-wavelength data set from the UV to FIR allows us to 
disentangle the SEDs into stellar, AGN, and star formation components, as 
described in an accompanying paper (Xu et al. 2015).

An important aspect of our study is the concentration on very luminous AGNs. 
The sample
is selected on the basis of AGN luminosity at 24 $\mu$m (which dominates over star formation
at this wavelength for every object). This band is a rough indicator of AGN luminosity 
at the typical redshift of z $\sim$ 1 \citep{spinoglio1995}, so the Type-1 
sample is complete for high luminosities (as a function of the appropriate redshift-dependent luminosity threshold).
Most of the sources have black holes of mass $\ge 10^8$ M$_\odot$, accreting at 3\% or
more of the Eddington rate. Because the sources in our study are emitting
close to the Eddington limit and hence uniformly represent approximately the maximum states of their
outputs, our conclusions are not hostage to modeling the AGN variability. 
There have been a number of suggestions that Type 2 AGN may lie in galaxies with 
relatively high rates of star formation, made both from
a theoretical perspective \citep{wada2002, ballantyne2006} and from observations \citep[e.g.,][]{maiolino1995, 
mouri2002, buchanan2006, deo2007, melendez2008, baum2010, castro2014, villarroel2014}. This result is controversial \citep[e.g.][]{pereira2010, diamond2012, merloni2014}; 
nonetheless, the Type-2 members of our sample test whether our conclusions are valid for such systems.

We find that there is substantial star formation  in most of 
the AGN hosts. However, we also show that the AGNs at $z<2$  reside in 
massive galaxies, and the majority of their hosts lie on or below the 
main sequence of normal star-forming galaxies. The high 
AGN and star forming luminosities need not have a direct causal connection, but may 
be linked through their mutual dependence on the masses of the host galaxies.


This paper is structured as follows. In Section 2
we summarize briefly the analyses of our sample in Xu et al. (2015). In Section 3 
we analyze the star-forming properties of the AGN host galaxies and discuss how they might 
influence AGN selection approaches. Section 4 is used to derive the expected 
joint behavior of star formation and black hole accretion luminosity. We then
compare this prediction with our results and with those in the literature.
In Section 5 we compare our results with theoretical studies of the co-evolution of 
massive galaxies and their central black holes. Our conclusions are summarized in Section 6. 
Throughout this paper we 
assume $\Omega_{M}$\,=\,0.3, $\Omega_{\Lambda}$\,=\,0.7, and $H_0$\,=\,70\,km\,s$^{-1}$\,Mpc$^{-1}$.

\section{Prior Results}

In Xu et al. (2015), we derived masses, black hole accretion rates, and star formation rates for 
the galaxies in this study. We also reached the following conclusions relevant to the following discussion:

\begin{enumerate}

  \item  About 50\% (107 out of 205) of the Type-1 AGNs 
in our sample are individually detected by \herschel. Among these 
AGNs, 68\% show  high levels of star formation (the star formation 
activity contributes over 50\% in the FIR).
\herschel\/ non-detected AGNs were studied using stacking analysis.
On average, they have a similar
level of AGN luminosity and similar optical colors, but the average star formation
activity is several times lower  compared with AGNs individually detected by \herschel.

\item Similarly, about 65\% (55 out of 85) of the Type-2 AGNs are individually detected by \herschel. However,
these objects tend to be at relatively low redshift and some of the 24 $\mu$m detections are a result of 
vigorous star formation, not just nuclear activity. We defined a High Luminosity Sample (HLS) and from it 
a Comparison Sample of 15 Type-2 AGN with
properties (M$_{BH}$, Eddington ratio, and redshift) that make them directly analogous to 
the Type-1 sample. 

\item The [OIII]$\lambda$5007 emission line, commonly taken to be a measure of Type-2 AGN luminosity, does
not show a 1:1 correspondence with this quantity in our data; instead, the line luminosity increases faster than the
bolometric AGN luminosity. 

\item  The FIR-detected Type-1 AGNs and the 15 matching Type-2 ones reside in massive galaxies ($\sim 1-2\times 10^{11} ~\rm  M_\odot$). They harbor supermassive black holes of $\sim  3 \times 10^{8} ~\rm  M_\odot$, 
which accrete at $\sim 10\%$ of the Eddington luminosity.


\item The 24 $\mu$m-selected sample of Type-1 AGNs includes about twice as many objects 
  as are identified through the Sloan Digital Sky Survey (SDSS), including the majority of the SDSS identifications. 
   The additional objects have redder optical colors than typical SDSS quasars, due to reddening 
   or intrinsically red quasar continua. There are also as many 24 $\mu$m-selected Type-1 AGNs 
as would be found in the X-ray in a survey to a similar bolometric luminosity limit. Therefore, our sample is
representative of powerful Type-1 AGNs in general (and of Type-2 AGNs with similar black hole masses); 
the infrared selection has not biased the sample toward some 
minority of AGNs particularly detectable in the infrared. 

\end{enumerate}

\noindent
We now build on the data and analysis in Xu et al. (2015) to investige the underlying relationship of the AGNs to their host galaxies.

\section{Host galaxy characteristics}

\subsection{Host galaxies lie on the main sequence}

The specific star formation rates (SSFRs: star formation rate divided by stellar mass) for the Type-1
sample are shown in Figure 1. Those based on stellar masses determined by spectral deconvolution and
the resulting estimate of the near-infrared stellar fluxes are shown as red filled circles. 
The values based on the indirect estimates of the galaxy stellar mass (based on the 
$M_*/M_{\bullet}$ relation and the $M_{\bullet}$ measurements for the Type-1 AGNs) are shown as blue filled squares.
The lower open blue squares show one side of the rms scatter  (Xu et al. 2015) in the relation for galaxies of the large masses typical
of our sample\footnote{The lower open squares also show the direct relation 
obtained by comparison of the K-band masses with those for
the same galaxies from the $M_*/M_{\bullet}$ relation. However, this value is anomalously low because the 
galaxies with stellar populations bright enough in the near infrared to outshine the AGNs and allow us to estimate 
$M_*$ will be biased within the scatter to have relatively large values of  $M_*/M_{\bullet}$.}. The upper 
open squares are a factor of four above the nominal values to accommodate both the scatter and a 
possible selection bias $--$ our very bright AGNs may have led us to host galaxies with relatively massive
black holes within the scatter of the $M_*/M_{\bullet}$ relation. See Xu et al. (2015) for further discussion.

As shown in Table 1, our full sample (Type-1 plus Type-2) consists of the maximally  
star-forming galaxies in very similar portions ($\sim$ 55\%) independent of redshift. 
Figure 1 shows the massive galaxy main sequence \citep{elbaz11}, an upper limit of three times
the MS SSFR (the rms scatter around the MS is 0.3 dex \citep{lutz14}, so this limit is 1.6 $\sigma$ high). We have 
adopted a lower  limit of 10\% because of evidence of bimodality in the SSFR, with a dividing 
line between active and inactive galaxies about an order of magnitude below the main sequence \citep{wetzel12}. 
The maximally star-forming quasar host galaxies generally fall on the main
sequence, with few galaxies outside these bounds. The remaining Type-1 AGN host galaxies will fall toward the bottom of the
main sequence zone in Figure 1, as shown by the stacking results, or if there are quiescent hosts,
below this zone  \citep{wetzel12}. 

The SSFRs of the Type-2 High Luminosity Sample (HLS) members as a function of redshift
are shown in Figure \ref{fig:stellar_mass} 
(along with similar results for the remainder of the Type-2 galaxies). 
The HLS members are indicated by squares, filled for the Comparison Sample, and the remaining
galaxies are small dots. 
The SSFRs of most of the galaxies are consistent with those of the 
main sequence of normal star-forming galaxies, although some are at the upper bound of this range.
The points for the Comparison Sample 
lie around the main sequence, showing that the members of the type-2 sample that most resemble 
the strongly-star-forming Type-1 sample similarly have main-sequence levels of star formation in their 
host galaxies. 

There are a small number of galaxies in the combined Type-1 and Type-2 sample that are undergoing
starbursts. The most noteworthy is J101805.93+385755.8, with a Type-2 AGN and a very strong component
due to young, hot stars in our decomposition. Because its stellar output appears to be significantly
influenced by young stars, the mass we have estimated is an upper limit, so its
SSFR in the figure is shown as a lower limit. However, starbursts also appear among field galaxies
\citep[e.g.][]{elbaz07}, within the small-sample statistics at about the same rate as in the AGN hosts.

\subsection{No strong dependence of host SSFRs on morphology}

For the Type-2 host galaxies, the AGN is sufficiently faint in the optical that we were able to 
examine the galaxy morphologies (Xu et al. 2015).  
Figure \ref{fig:morphology} shows the  IR star formation luminosity and SSFR as a function 
of redshift for different types of host galaxy. 
We do not find any correlation between the star formation and 
the morphological types. The five AGN hosts showing probable interaction do not have more 
star formation than the other types, as found for a much larger sample by \citet{villforth14} and \citet{sabater15}, among others. 
All types of hosts form stars at rates consistent with normal star forming galaxies.  
However, our study of morphologies is limited by the resolution of the images and the small sample size.

\subsection{Use of ${\mbox{\rm \oiii}}$ as a AGN Luminosity Indicator}

Xu et al. (2015) found a significant departure from the expected 1:1 relation between ${\mbox{\rm \oiii}}$ and bolometric AGN luminosity in the direction of an increasing [OIII] luminosity for more luminous AGN, an effect also reported by 
\citet{lamassa10},  Hainline et al. (2013), and Shao et al. (2013) (in the latter case for only one of two infrared bands). 
Hainline et al. (2013) propose that the effect arises because of limitations on the extent of the narrow line region with
increasing AGN luminosity. The fact that we have an independent estimate of the star formation in the host galaxy lets us 
identify an additional possible contributor.  
Conventional groundbased spectra (fiber diameters or slit widths of 1 - 2 arcsec) of galaxies at moderate to high redshift and with active nuclei might return ${\mbox{\rm \oiii}}$ measurements contaminated by star formation in the host galaxies. 
For example, our 1\farcs5 fibers subtend 6.7 kpc at z = 0.3 and 10 kpc at z = 0.6. 
Since in typical AGN samples the most luminous members tend to be 
the most distant, this would yield an increase in [OIII] more rapidly than just in proportion to the AGN luminosity. 

We can evaluate this possibility quantitatively. 
From the relations in Kennicutt (1998), 
the H$\beta$ luminosity in a star forming galaxy is about 0.2 $\times$ A$_{H\alpha - H\beta}$\% of the infrared luminosity (assuming Case B and an extinction of A$_{H\alpha - H\beta}$ between H$\alpha$ and H$\beta$). From Moustakas \& 
Kennicutt (2006), the [OIII]$\lambda$5007 line is on average roughly the same intensity as H$\beta$ in the integrated
light of star-forming galaxies. Therefore, assuming modest extinction, we can adopt L([OIII)]/L(IR) $\sim$ 0.1\% for typical  
star-forming galaxies.  From Diamond-Stanic et al. (2009) and Rigby et al. (2009) one finds that the luminosity of [OIII]$\lambda$5007 is about 0.1\% of the bolometric luminosity of a Type-2 AGN. This value is consistent with the 0.17\% for the extinction-free [OIII]$\lambda$5007 luminosity as adopted by \citet{heckman14}. Comparing these values with the relative luminosities in Table 8 of Xu et al. (2015) indicates that there is indeed a possibility of contamination of the [OIII] line by emission due to star formation for spectrograph fibers that include a substantial part of the host galaxy. 
One might expect from this hypothesis that
the ratio of [OIII] to H$\beta$ would tend to decrease with increasing redshift. Our data show a weak tendency in this direction, 
but the scatter is large and the result is not statistically significant; it should be tested in larger samples.

\section{Connection between Star Formation and Active Nucleus Luminosity}

We have just shown that 
the strongly star-forming, \herschel-detected, AGN hosts nearly all lie on the star-forming  galaxy main sequence.  
Our stacking analysis indicates that the \herschel-non-detected galaxies lie in 
the lower part of the main sequence, although some could also lie below it (Xu et al. 2015). 
The morphology study is also consistent with this result --
most of the AGNs in our sample are not interacting, and hence not interaction-induced starbursts. 
Even those that are interacting do not have particularly high levels of star formation.
These results indicate there is no obvious causal relation between nuclear activity and elevated 
star formation in a host galaxy; we probe this possibility in more depth in this section by examining 
the nature of the apparent connection between star formation and AGNs within individual galaxies. 

\subsection{The predicted $L_{\rm SFstrong} \propto L_{\rm AGN}$ relation}

A general correlation between star formation rate, i.e. infrared luminosity, and AGN luminosity for quasars
was found some time ago,  (e.g., Lutz et al. (2008), Netzer(2009), Shao et al. (2010)).
There have been suggestions that the correlation implies a causal connection between the two kinds
of activity. To evaluate these suggestions, we first derive the connection between the star forming
luminosity and that of the AGN assuming the AGN hosts are normal galaxies, forming
stars sufficiently vigorously to place them on the main sequence. To distinguish them from 
galaxies at the lower part of the main sequence or that are quiescent, we designate them as 
having {\it Herschel} fluxes indicative of strong star formation. 
For the purpose of discussion the derivation that follows makes the assumption that 
their level of star formation is {\it not} necessarily affected by the presence of an AGN.

The output of a quasar powered by a black hole with mass $M_\bullet$ is 
typically expressed in terms of the Eddington limit, i.e.,
\begin{equation} 
L_{\rm AGN} = \epsilon  L_{E}= 3.2\times 10^4  \epsilon \left(\frac{M_\bullet}{\rm  M_\odot} \right) \rm L_\odot,
\label{eq:qso_eddington}
\end{equation}
where $\epsilon$ ($\sim 0.1$ for luminous AGNs) is 
the efficiency of converting gravitational potential energy to electromagnetic radiation.
Conversely, Equation \ref{eq:qso_eddington} can be used to estimate $M_\bullet$ for a
quasar near maximum output. 
The black hole mass is related to the stellar mass of the quasar host galaxy by 
$M _{*}/M_\bullet = \eta \sim 700$. 
The SFR for galaxies on the main sequence is 
\begin{equation} 
\rm {SFR} = C_1   M_{*}^{\beta} 
\label{eq:sfr_mstar}
\end{equation}
Various values of $\beta$ have been derived, all consistent with 0.8. 
Elbaz et al. (2007) find $\beta =0.9$ for their data and also show that the Millenium model gives $\beta =0.8$; Noeske et al.
(2007) derive $\beta=0.7$; Tyler et al. (2013) find $\beta = 0.71$. 

Equation \ref{eq:sfr_mstar} applies locally. Star formation in galaxies can be fitted by
luminosity evolution as $(1 + z)^\alpha$ with $\alpha \sim 3.2$ 
for $0 < z < 1 $ (Rujopakarn et al. 2010), and with a smaller value
 of $\alpha$ at higher z.
We also have Equation \ref{eq:sfr_lfir_eq},

\begin{equation} \frac{{\rm SFR}}{{\rm  \rm M_\odot~yr^{-1}}} =
  1.2\times10^{-10}\left(\frac{L_{\rm SF,IR}}{{\rm \rm L_\odot}}\right).
\label{eq:sfr_lfir_eq}
\end{equation}

\noindent
the classic relation between  infrared luminosity (from
8 to 1000 $\mu$m) and the SFR (Kennicutt 1998)\footnote{adjusted to our adopted IMF - see Xu et al. (2015)}.
Combining Equations  \ref{eq:qso_eddington},  \ref{eq:sfr_mstar}, \ref{eq:sfr_lfir_eq}, 
and the $M _{*}/M_\bullet$ correlation, we get

\begin{equation} 
L_{\rm SF} =C_2  (1+z)^\alpha \left(\frac{\eta L_{\rm AGN}}{\epsilon} \right)^\beta.
\label{eq:lsf_lagn}
\end{equation}
This is consistent with Netzer's result (2009) of $L_{\rm SF} \propto L_{\rm AGN}^{0.8}$.
The quasar tracks in Lutz et al. (2008)  and 
Shao et al. (2010)  can also be fitted well by the correlation
$L_{\rm SF} \propto L_{\rm AGN}^{\beta}$ with $\beta$ $\sim$ 0.8. Our derivation of  Equation \ref{eq:lsf_lagn}
shows that this relation can arise through the dependence of {\it both} the SFR {\it and} the mass of the
SMBH on the mass of the host galaxy, without any other connection between the two activity types.

\subsection{The observed correlation between AGN and SF luminosity\label{agn_sf}}

\subsubsection{Approach}

We now use our own sample to gain more understanding of the relation of AGN and star forming 
luminosities reflected in the diagrams of Lutz et al. (2008), Netzer(2009), and Shao et al. (2010). 
To compare with the prediction derived in the preceding section, we 
will determine the slope of the $L_{\rm SFstrong} \propto L_{\rm AGN}$ relation. We have
added ``strong" to the SF subscript as a reminder that we are not interested in the slope for all of 
the galaxies in the AGN sample, but only for those toward the upper range of star formation and hence that
should fall on or above the galaxy main sequence. 

The dependence of the
detectable luminosity on z for our flux-limited sample might be expected to result in strong selection effects, leading to
a slope that was influenced by the change in detection rate with {\it Herschel} as a function of redshift.
Fortunately, this concern is unfounded. As shown in Table 1, the combination of the approximate
proportionality of the SFR and AGN luminosity, plus the negative K-correction in the far infrared,
have resulted in our overall sample (Type 1 plus Comparison Sample of Type 2 AGN) to be
detected by {\it Herschel} at the same rate independent of redshift. Therefore, 
we can calculate the slope on the basis of the {\it Herschel}-detected galaxies knowing that
they are in the upper 55$^{th}$ percentile of star formation irrespective of redshift. 

Inclusion of galaxies with lower levels of far-IR emission 
could, in fact, bias the result because the undetected part of our sample may include 
quiescent host galaxies. For example,
a proportion of $\sim$ 19\% quiescent galaxies, as found in the PG sample \citep[e.g.,][]{shi14} would, if included in
the slope calculation, make the overall slope flatter than if it were based only on the 
strongly star forming part of the sample. We have simulated this situation using a model with 81\% of the 
simulated galaxies having SFRs strictly proportional to AGN luminosity and 19\% with SFRs
equal to each other and independent of AGN luminosity. A linear fit (in log-log space) has a slope of 0.73. 
Thus, one might conclude despite a 1:1 relation between
AGN luminosity and SFR for the strongly star forming galaxies, that the intrinsic relation was not 1:1.
For this reason, we do not include the stacked results in fitting the behavior, nor do we use
statistical methods to extend into the range where we do not have solid detections.


\subsubsection{Results}

In  Figure \ref{fig:agn_sf} we combine the
Type-1 sample and Type-2 Comparison Sample; the AGN and SF luminosities cover three and 2.5  orders of magnitude for
the whole AGN sample, respectively.  The best fit is $L_{\rm SF} \propto L_{\rm AGN}^{0.59} $. We
have tested whether a simple linear regression is an adequate fit by fitting with polynomials of
order 2, 3, and 4; none of them had any significant effect on the reduced $\chi^2$, so by Occam's Razor
the linear fit is preferred. The Jarque-Bera test indicates that the residuals from the linear fit
are normally distributed (p = 85.1\%), in agreement with the Shapiro-Wilk Test (p = 39.2\%). A fit
to the residuals indicates a rms scatter of $\sim$ 0.4 dex, which is consistent with expectations from
the scatter of $\sim$ 0.3 dex for purely star forming galaxies around the main 
sequence \citep{brinchmann04, lutz14} with a modest increase due
to the expected scatter in Eddington ratios. Various alternative samples, e.g. discarding galaxies 
where the warm component fit could account for significant fraction
of the far infrared luminosity, do not change the slope significantly.

We now consider whether we can rule out the hypothesis that the star formation in the host galaxies has a 1:1 
correlation with the AGN luminosity, as might be expected if they are linked causally. To do so, we use the $\Delta \chi^2$ method 
\citep{press2002} to evaluate the range of slopes that can fit the data in  Figure \ref{fig:agn_sf}. 
Because the errors on the luminosities are not known {\it a priori} , we set them to
be equal and to a value so the reduced $\chi^2$ = 1 for the best fit.  We then computed 
$\chi^2$ for other slopes and derived the corresponding probability that this parameter 
could be as large or larger than the computed value by chance. The result is that slopes of 0.7 are
consistent with our measurements at $\sim$ 37\% probability, slopes of 0.8 at $\sim$ 15\%, 
slopes of 0.9 at $\sim$ 1.7\% probability, and slopes of 1.0 are consistent only at the $\sim$ 0.02\% level.  
That is, the observations do not appear to support the expectation of a 1:1 correlation 
if there were a causal connection star forming and nuclear activity, but instead suggest that 
the observed relation could arise only due to their mutual
dependence on galaxy mass.

As shown in Figure~\ref{fig:agn_sf}, there is a strong trend of increasing 
luminosity with redshift as a natural consequence of using a pencil 
beam survey to select the sample, so the relative volume in the survey increases rapidly with increasing redshift. 
However, this could also mean that the trends of star formation and AGN activity are dependent on evolution in ways not 
included in the analysis above. There are two obvious possibilities. First, the SMBHs of the particular maximally luminous AGNs 
in our study might grow substantially between z $\sim$ 2 and z $\sim$ 0.5 (although on average it is believed that the growth of the most massive black holes is 
nearly complete by z = 2 \citep[e.g.,][]{marconi04,trakhtenbrot12}). In this case, the Eddington limits will go up and the 
most luminous AGNs will be more luminous locally, moving the low redshift/low luminosity points in  Figure~\ref{fig:agn_sf} 
to the right. The second is that the main sequence level of specific star formation 
decreases as we approach the current epoch, by about a factor of 3-4 between z $\sim$ 1.5 and 0.5 \citep{elbaz11}. 
This behavior would move the  low redshift/low luminosity points in  Figure~\ref{fig:agn_sf} 
down. Both effects would therefore make the observed slope steeper than it would be without 
these potential types of evolution. That is, the fitted slope is an 
upper limit at least so far as these two first-order evolutionary terms are concerned. 
Although this does not rule out the possibility of other, more complex changes affecting the 
slope, the existing evidence strongly suggests that the 1:1 relation does not hold.

As a final test, we have analyzed the behavior of the Eddington Ratio vs. the SSFR (using only values of the
SSFR based on the photometric mass determinations). Since these quantities are 
normalized to remove slopes to first order, this comparison should show no dependence if there is no
link between the quantities. We find that the slope of the relation is equal to $0.02 \pm 0.30$. The large 
uncertainty arises because of the limited range of the Eddington ratios and the large scatter of both 
quantities. 

\subsection{Previous Studies of the Connection between AGN and SF}

\subsubsection{The Relation between SF and AGN Luminosities}

There are several previous studies of the relation between AGN and SF luminosities. 
A strong correlation, $L_{\rm SFstrong} \propto L_{\rm AGN}^{0.8} $,  
has been demonstrated over more than five orders of magnitude in luminosity \citep{netzer09}.  
The sample included PG quasars at $z\sim 0.1$ (Schweitzer et al. 2006), 
AGN-dominated Type-2 galaxies at low redshift ($z<0.25$) 
identified using SDSS spectra,  
and strong submillimeter quasars at $z\sim 2-3$ (Lutz et al. 2008). 
These studies use different SF and AGN indicators for different types of AGN/host galaxy 
sources, and use data from surveys with different depths. 
Nonetheless, a correlation between AGN and SF luminosities 
was found for the luminous AGN systems within the whole sample.
This study led to suggestions of direct links between black hole and bulge growth, at
least for relatively luminous AGN.
The most luminous AGNs in these studies should be similar to our AGN sample (either Type 1 or 
Type 2), i.e., accreting at $\sim$ 10\% of Eddington. However, as noted in the
preceding section, the coefficient in this correlation is in good agreement with the value 
expected if the correlation arises only because of the mutual dependence of 
the rate of star formation and of $M_{\bullet}$ on galaxy mass. 

For X-ray-selected, low or moderate  luminosity  AGNs at $z\sim 0-3$,  there is no 
strong correlation between AGN and SF luminosities
for individually  detected AGNs  (e.g., Shao et al. 2010; Rosario et al. 2012; 
 Mullaney et al. 2012; Rosario et al. 2013), 
even though these AGN host galaxies lie on the galaxy main sequence (Mullaney et al. 2012). 
One possible reason is that low or moderate  luminosity AGNs have a large scatter
in the distribution of Eddington ratios (e.g., Lusso et al. 2012). 
This scatter will also affect the correlation
between AGN and SF luminosities. 
A simple model of the scatter in $L_{\rm X}$ due to AGN variability
indicates that the correlation also exists for low and moderate luminosity AGNs
(e.g.  Chen et al. 2013;  Hickox et al. 2013).

The correlation may be more evident for higher luminosity AGNs. 
Rafferty et al. (2011) studied the far infrared properties of AGNs in fields 
with very deep X-ray survey data, finding a strong correlation between star-forming 
luminosity and the presence of powerful AGNs. However, they point out that this 
effect could arise through the mutual dependence of both phenomena on stellar mass. 
Rosario et al. (2012) worked with a sample of X-ray selected AGNs using PACS data to 
study the correlation between $L_{\rm SF}$ and $L_{\rm AGN}$. 
They found a strong correlation between $L_{60 \rm \mu m}$ and $L_{\rm AGN}$ for high luminosity AGNs
 ($L_{\rm AGN} \sim 10^{44}-10^{46} \rm ~ergs~s^{-1}$) at 
$z<1$,  but not at high redshift ($z>1$) (but with a very small subsample, only 4 objects). 
They claimed  the reason for this trend is that at $z<1$
 major-mergers play important roles at the high luminosity end of both AGN and
star formation activities, and the importance of major-mergers decreases at $z>1$.
However, assuming the Type-2 AGNs are representative of the $z<1$ category, we find few examples of mergers, 
contrary to their prediction. Our result agrees with those from much larger samples \citep{villforth14,sabater15}.

Rovilos et al. (2012) report that a sample of X-ray selected AGNs have elevated SSFRs.  
In their Figure 9, 8 out of 13 X-ray quasars with intrinsic 
$L_{2-10 \rm ~keV }> 10^{44} \rm~ergs~s^{-1} $ at $z<2$ are above the galaxy main sequence, 
and only 5  are on the galaxy main sequence or
reside in  quiescent galaxies. 
However, they may use the wrong AGN templates to decompose the SEDs. 
By matching the X-ray AGN source catalog for the CDFS, we are able to identify 
two sources at $z=1.031$ and $ z=1.216$ in Fig 2 of Rovilos et al. (2012). 
The hardness ratios of these two sources are  -0.16 and 0.45, respectively, using the data from the 
Chandra 2Ms survey catalog (Luo et al. 2008).  
This puts the first source in the unobscured AGN (Compton thin) category. 
The second is in a transitional region, but it has also been found to be Compton-thin (Tozzi et al. 2006).
Therefore, the sources should not be fitted with an obscured template as was done by Rovilos et al. (2012).
In particular, a Type-1 AGN template should be used to  fit the UV emission rather than a UV-bright galaxy template.  
Therefore the stellar masses,  and hence the specific SFRs of these two X-ray quasars
are not reliable. 
In general, fitting the UV with a young stellar population to compensate 
for the lack of UV output from the absorbed AGN template will 
significantly underestimate the stellar mass and overestimate the SSFR. 

\subsubsection{The Coincidence of Star Forming and Nuclear Activity in the Same Galaxy}

Although we have argued that there is no compelling evidence for
a causal link between elevated star formation and nuclear activity, a 
possible counter-example is provided by \citet{symeonidis13}.
They studied  a sample of IR-luminous galaxies (detected at 70 $\mu$m), 
X-ray detected Type 2 AGNs, and hybrid AGN/SF sources from COSMOS.
They found that the IR luminous/star-forming galaxies 
and Type 2 AGN hosts have a significant overlap in 
color-magnitude space ($U-V/M_V$) and color-color space ($U-V/V-J$),
thus minimizing selection effects between the samples. 
They calculated the predicted numbers of hybrid AGN/SF sources
in 0.5 magnitude bins of $M_{\rm V}$ and 0.25 magnitude bins in $U-V$ and $V-J$ 
assuming that the presence of an AGN and the star formation-powered IR emission  
in a given galaxy are independent events.
They found that the predicted numbers of hybrid AGN/SF sources under 
this assumption
are two-four times lower than the observed numbers; the discrepancy
is removed if the two activities are not completely independent.
They interpret this behavior as evidence for a causal link between black hole 
accretion and star formation. 

The approach of \citet{symeonidis13} depends on there being no underlying
selection bias that might favor {\it both} high rates of star formation {\it and} luminous nuclear
activity. Since the detection rates in \citet{symeonidis13} are low ($\sim$ 0.6\% for infrared-emitters
and $\sim$ 0.3\% for AGN), the detected galaxies are at the extreme
high end of the luminosity distributions and any biases can have a strong effect that might
mimic a causal connection. 
Both AGN and star forming luminosities have a dependence on 
host galaxy stellar mass. Star formation rates generally go as the 0.8 power of the mass
(as already discussed). Although this value is strictly for the slope
of the main sequence, the envelope for the star formation rates in galaxies 
above the main sequence has a similar slope (e.g. \citet{wuyts11}, Figure 1, central panel)
and in any case the most luminous infrared galaxies, which will be favored
for detection in \citet{symeonidis13}, tend to be very massive \citep{rothberg13} and 
hence at the extreme high tail of the star formation rate distribution.  
For an AGN, the SMBH mass, and hence the Eddington Luminosity, is correlated with the stellar mass in
the bulge. The most luminous AGNs will tend to have the highest Eddington 
Luminosities and thus massive galaxies are favored for detection. 

These mass dependencies undermine the independence of high AGN and star-forming
luminosities.   The sample selected by \citet{symeonidis13} on the basis of color-color
behavior does not control directly for mass and hence is particularly susceptible
to this issue. We therefore examine in more depth their other selection approach, 
on the basis of $U-V$ and $M_V$. Although the host galaxy masses correlate with galaxy optical 
properties,  e.g. $M_V$, the correlation between $V$-band magnitude and the stellar mass 
has a large scatter, of order a factor of 10 in a $\Delta M_{\rm V} \sim  0.5$ magnitude bin
(e.g., Shapley et al. 2001; Savaglio et al. 2005; Kannappan et al. 2007). 
Therefore, it is possible that the apparent correlation of AGN and 70 $\mu$m sources 
is influenced by their dependence on mass even in a sample selected in 0.5 magnitude 
bins of $M_{\rm V}$. 

We ran a simple simulation to test this possibility. 
In each bin of $M_{\rm V}$, within a bin size of 0.5 mag, say [$M_{\rm V,0}$, $M_{\rm V,0}+0.5$],
we assume the mass to be evenly distributed in the range of [$1M_{*,0}$, $10M_{*,0}$]. 
Specifically, we distribute the mass over 10 small bins in this range to compare 
with the case where $M_{\rm V}$ is a perfect mass indicator and predicts host stellar mass tightly correlated 
with $M_{\rm V}$, [$4M_{*,0}$, $5M_{*,0}$]. 
We use the 
SFR probability distribution around the main sequence from \citet{brinchmann04}  to estimate the  probability distribution of 
the IR emission as a function of host galaxy stellar mass; this distribution includes starbursting galaxies that might dominate the 
far infrared detections. 
For luminous AGNs, we 
assume the Eddington ratio probability distribution follows the G-model in 
Shankar et al. (2013).\footnote{a Gaussian in log $\lambda_c$ with dispersion of 0.3 dex, and centered
at log $\lambda_c = -0.6$.} For each given galaxy stellar mass, 
we estimate the black hole mass using the $M_*/M_\bullet$ relation.
Then we convert the Eddington ratio probability distribution
to an AGN luminosity probability distribution. 

We use these two distributions of IR and AGN luminosity 
at a given mass to estimate the probability that the IR or 
AGN luminosity will fall above the selection threshold for 
the cases: 1.) where the stellar masses all lie within 
a narrow range; or 2.) where the stellar masses are 
spread over a range of a factor of ten. 
The results for the case of AGNs are shown in Figure \ref{fig:lagn_pro}; those for IR luminosity are very similar.
We found that Case 2  has a much larger probability that the IR or AGN 
luminosity falls above the detection/selection threshold, and that the increased 
detection rate is largely due to the most massive galaxies in the assumed mass range. 
For example,  $M_{\rm V}$ in the range of [-20.5, -21] magnitude
corresponds (with mass errors) to a range of $4 \times 10^9 - 4 \times 10^{10}~ \rm M_\odot$.
If the stellar masses are all in the center of this range, 
the probability of sources with AGN X-ray luminosity above $6\times 10^{42} $ \ergs~ is 0.3\%, 
as in \citet{symeonidis13} (M. Symeonidis, private communication). This value is also typical
of the detection rates in their full sample.  
If the stellar masses spread out over the 10-times larger range, 
the probability of the sources of this $M_V$ with AGN X-ray luminosity above $6\times 10^{42} $ \ergs~ is 
six times higher.
We obtain a similar result for IR luminosities.

Our conclusion is that the galaxy stellar masses strongly affect the 
detection rates for SFR at 70 $\mu$m and for AGN.  
This behavior will reproduce the tendency for the two phenomena to occur together, as observed by \citet{symeonidis13}, 
from which they inferred a causal connection between star formation and nuclear activity. 
However, in our simulation, the behavior arises purely because both phenomena 
have a strong mass dependence; there may be no need to hypothesize any additional
causal connection.

\section{Implications of not finding a direct link between AGNs and star formation in their host galaxies}

\subsection{Possible lack of a connection between star formation and black hole accretion}

We have shown that the AGN host galaxies have normal, main-sequence levels of star formation. 
Although high levels of star formation and high AGN luminosities tend to occur together, the 
luminosities from these two phenomena do not correlate 1:1 as might be expected if there were 
a causal relation between the two activity types. One possible conclusion is that there is, in fact, no connection 
within the range of cosmic epochs probed in this study. 
However, if the SFR and AGN were completely decoupled, 
then there might be too much scatter in the $M_*/M_\bullet$ relation, 
even if their growth somehow averaged out over time. One proposal is that the 
correlation is a result of the process of hierarchical assembly in a 
$\Lambda$CDM Universe through the Central Limit Theorem \citep{peng07,jahnke11}, i.e., the link is causal
but was established early in the assembly of these galaxies.

\subsection{Is black hole accretion delayed relative to star formation?}

Another possibility is that any correlation is not simultaneous, but  the AGN 
activity is delayed. The timescales of the starbursts in ULIRGs are estimated to be short, 
that is no more than  $10^7$ to $10^8$ years in the most luminous systems, 
and perhaps in some cases as small as $5-10 \times 10^6$ years  
(Genzel et al. 1998, Thornley et al. 2000).  These short timescales 
are in agreement with theoretical models (e.g., Chakrabarti et al. 2008; 
Hopkins et al. 2008; Cen 2012; Hopkins et al. 2013; Hayward et al. 2013). 
The models for AGN luminosity show similarly short peaks, but there is a 
range of predictions on when they occur relative to the peak of the 
star formation. Bekki et al. (2006) model the two phases to be nearly 
simultaneous. Hopkins et al. (2008) and Hopkins (2012) predict 
a delay of order 80 Myr for the peak of AGN luminosity, 
but with a substantial overlap when both the AGN and starburst are very luminous. 
Cen (2012) argues for a longer delay, with the peak of AGN activity after 
the prime starburst has died out. There is already some observational evidence that
the primary host galaxy star formation may occur up to $10^9$ years 
before the peak AGN luminosity \citep{wild07}.

Our sample of AGN is selected in the mid-IR 
and hence is optimized for the early, blow-out, infrared-luminous  
phase in quasar development (e.g., Hopkins et al. 2008; 
Georgakakis  et al. 2009; Glikman et al. 2012). Thus, any delays between 
the peak of star formation and the peak of AGN infrared luminosity should 
be minimized compared to the delays for  AGN total luminosity or X-ray luminosity, for example.
Nonetheless, we find little evidence for simultaneity in elevated 
star formation and near-Eddington AGN activity. 
This result would favor models with relatively long delays, such those of Wild et al. (2010) and Cen (2012).


\section{Summary}

We studied the properties of a 24 $\mu$m-selected spectroscopically-identified 
AGN sample using a
multi-wavelength dataset from  \galex, SDSS, UKIRT, WISE,
\spitzer/MIPS, and \herschel. 
We summarize the results from this study (including results from the accompanying paper by Xu et al. (2015)) as follows.
\begin{enumerate}

\item As also found, e.g., by Hainline et al. (2013), the strength of the [OIII]$\lambda$5007 line 
increases more rapidly than proportionately to bolometric AGN luminosity (Xu et al. 2015). At relatively high redshift (and 
hence high AGN luminosity), detection of [OIII] 
emission from parts of the host galaxy within the spectrograph fiber may contribute to this effect.

  \item A warm excess in the MIR was found for eight Type-1 AGNs compared with a local quasar template (Xu et al. 2015). 
	This warm excess is more prominent at higher redshifts within our sample. 
        It is not clear whether this change is due to evolution, 
	or whether the warm excess is confined to very luminous quasars.  

\item For z $>$ 0.3, selection at 24 $\mu$m yields a sample of highly luminous Type 2 AGN (Xu et al. 2015).
 At lower redshifts, the sample is significantly contaminated by star forming galaxies.

\item  For $0.3 < z < 0.8$ the numbers of luminous Type 1 and Type 2 AGN are similar (Xu et al. (2015)).

\item The host galaxies of luminous Type-1 and Type-2 AGNs have SSFRs consistent
 with the galaxy main sequence trend, i.e. they are normal, star forming galaxies. 
  
\item  There is a strong correlation between the IR luminosity of the star formation component and 
   the AGN total luminosities.  

  \item However, the correlation differs significantly from the 1:1 correspondence 
that might be expected if the star formation and AGNs were related directly, and it could arise just because 
   the BH mass (and hence Eddington limit) and the star formation 
   are both correlated with the galaxy mass, rather than  requiring 
   a causal connection between the star formation and the nuclear activity.

\item Although a number of evolutionary models indicate that peaks of
star formation and nuclear activity should occur closely in time, our results 
either indicate no connection in the two forms of activity over the cosmic 
epochs probed by our sample ($0.5 < z < 2$), or
favor models where fueling of an AGN is well 
removed in time from the triggering of elevated star formation in 
the host galaxy.

\end{enumerate}

\section{Acknowledgements}

We thank Xiaohui Fan, Desika Narayanan, and Dan Stark for helpful discussions.
Jacopo Fritz provided AGN SED templates. We also thank
Yong Shi for communicating results on quasar aromatic band measurements in advance of publication.
This work is based in part on observations made with {\it Herschel}, a European Space Agency Cornerstone Mission with 
significant participation by NASA. Additional observations were obtained with {\it Spitzer}, operated by JPL/Caltech. We acknowledge NASA funding for this project through an award for research with {\it Herschel} issued by JPL/Caltech. CPH was funded by CONICYT Anillo project ACT-1122. GPS acknowledges support from the Royal Society. Additional support was provided through contract 1255094 from JPL/Caltech to the University of Arizona. This paper also is based in part on work supported by the National Science Foundation under Grant No. 1211349.

Funding for the SDSS and SDSS-II has been provided by the Alfred P. Sloan Foundation, the Participating Institutions, the National Science Foundation, the U.S. Department of Energy, the National Aeronautics and Space Administration, the Japanese Monbukagakusho, the Max Planck Society, and the Higher Education Funding Council for England. The SDSS Web Site is http://www.sdss.org/.

The SDSS is managed by the Astrophysical Research Consortium for the Participating Institutions. The Participating Institutions are the American Museum of Natural History, Astrophysical Institute Potsdam, University of Basel, University of Cambridge, Case Western Reserve University, University of Chicago, Drexel University, Fermilab, the Institute for Advanced Study, the Japan Participation Group, Johns Hopkins University, the Joint Institute for Nuclear Astrophysics, the Kavli Institute for Particle Astrophysics and Cosmology, the Korean Scientist Group, the Chinese Academy of Sciences (LAMOST), Los Alamos National Laboratory, the Max-Planck-Institute for Astronomy (MPIA), the Max-Planck-Institute for Astrophysics (MPA), New Mexico State University, Ohio State University, University of Pittsburgh, University of Portsmouth, Princeton University, the United States Naval Observatory, and the University of Washington.

This publication makes use of data products from the Wide-field Infrared Survey Explorer, which is a joint project of the University of California, Los Angeles, and the Jet Propulsion Laboratory/California Institute of Technology, funded by the National Aeronautics and Space Administration.

\clearpage

\begin{deluxetable}{lcccccc}
\tabletypesize{\footnotesize}
\tablecolumns{7}
\tablewidth{0pt}
\tablecaption{Fraction of Herschel Detections in Full Sample}
\tablehead{\colhead{}                   &
           \colhead{total}                    &
           \colhead{$0 \le z < 0.5$}     &
           \colhead{$0.3 \le z < 0.5$}     &
           \colhead{$0.5 \le z < 1$}          &
           \colhead{$1 \le z < 1.5$}         &
           \colhead{$1.5 \le z < 3$}         }
\startdata
{\it Herschel}-detected	&   157   &	55			&   24    &  	38		&  29			&   35			\\
total                          	&   285   &	96	 		&   51   &	76 		&  54			&   59			\\
fraction detected		&   0.55 &	$0.57 \pm 0.08$ 	&   $0.47 \pm 0.10$ &  $0.50 \pm 0.08$&  $0.54 \pm 0.10$	&  $0.59 \pm 0.10$	\\

\enddata

\end{deluxetable}

\begin{figure}[!hbt]
\centering
\includegraphics[width=0.8\textwidth,angle=0]{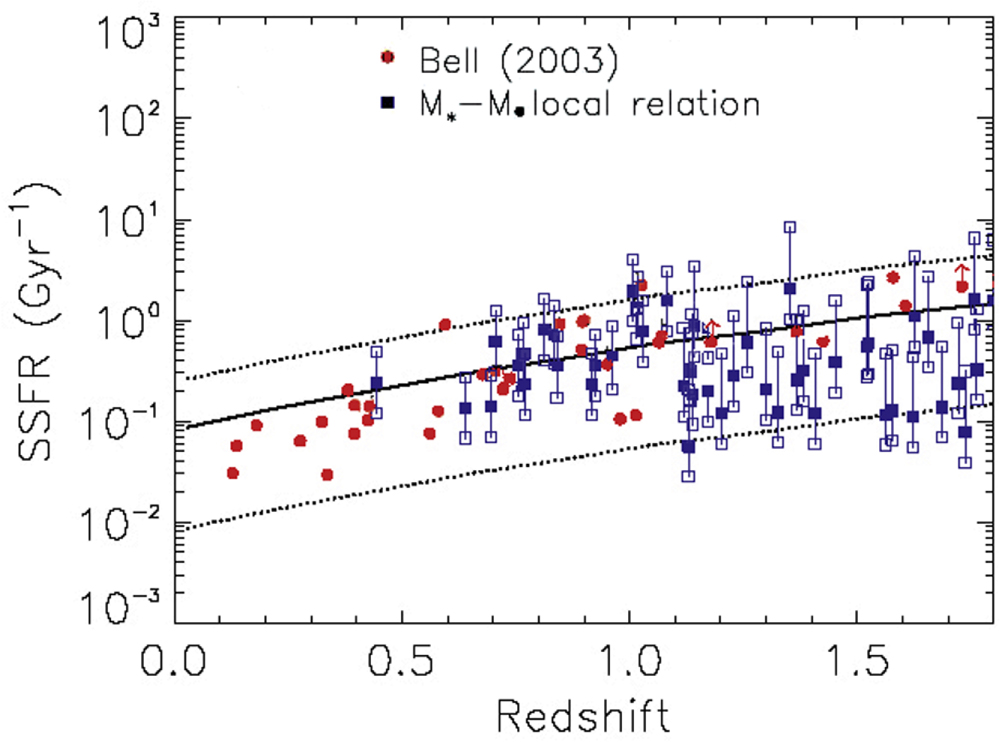}
\caption{SSFR versus z for {\it Herschel}-detected Type-1 AGNs. 
The solid line is the SSFR of main sequence galaxies (Elbaz, et al. 2011).
The dotted lines are a factor of three above or 10 below the SSFR of main sequence galaxies, respectively.
The filled red circles represent sources with stellar mass estimated  by K-band luminosity. 
The filled blue squares are represent sources with stellar mass estimated from the local mass ratio $M_*/M_{\bullet}= 700$. 
The lower open squares are the SSFR values using masses a factor of two lower than the local mass ratio, representing the 1-$\sigma$ scatter in the $M_*/M_{\bullet}$ relation. The upper open squares are show the upper rms 
scatter for z $<$ 1.1 but above that value are a factor of four above the nominal
estimates, allowing for an additional possible factor of two systematic error in the $M_*$ estimates due to 
possible evolution and/or selection of 
the galaxies on the basis of exceptionally luminous AGNs}
\label{fig:stellar_mass_more}
\end{figure}

\begin{figure}[!hbt]
\centering
\includegraphics[width=0.8\textwidth, angle =0]{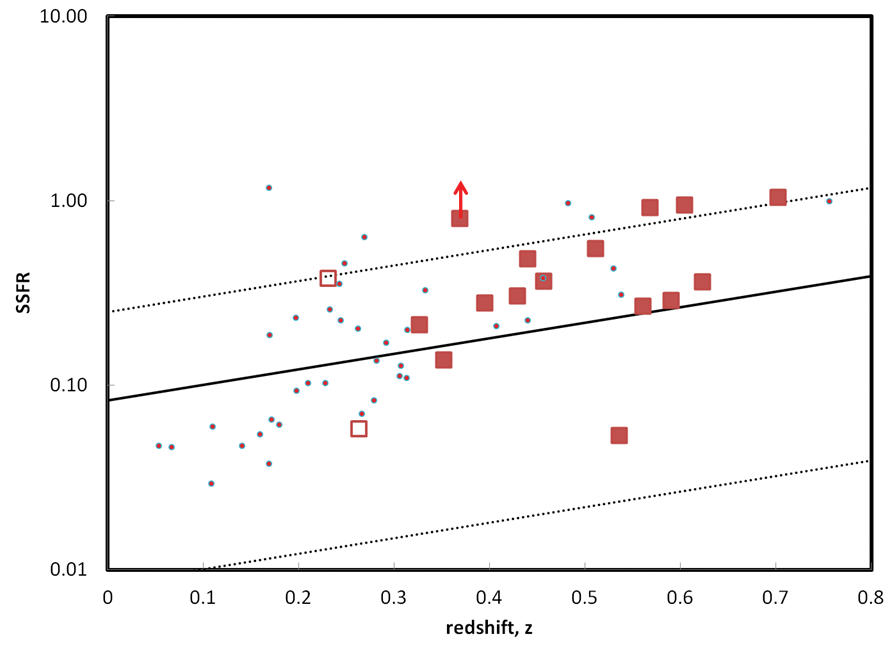}
\caption{SSFR versus redshift for the \herschel-detected Type-2 
AGNs. The solid line is the SSFR of main 
sequence galaxies (Elbaz, et al. 2011).
The dotted lines are  a factor of three times or  10\% of the SSFR of main sequence galaxies, respectively.
The filled boxes are for our Comparison Sample, while 
the unfilled boxes are the two sources with similar AGN properties but at z $<$ 0.3. The small dots
are for the remainder of the {\it Herschel}-detected type 2 galaxies. The lower
limit is for J101805.93+385755.8, which is undergoing a strong starburst and for which our
mass estimate is an upper limit.}
\label{fig:stellar_mass}
\end{figure}

\begin{figure}[!hbt]
\centering
\includegraphics[width=0.9\textwidth, angle =00]{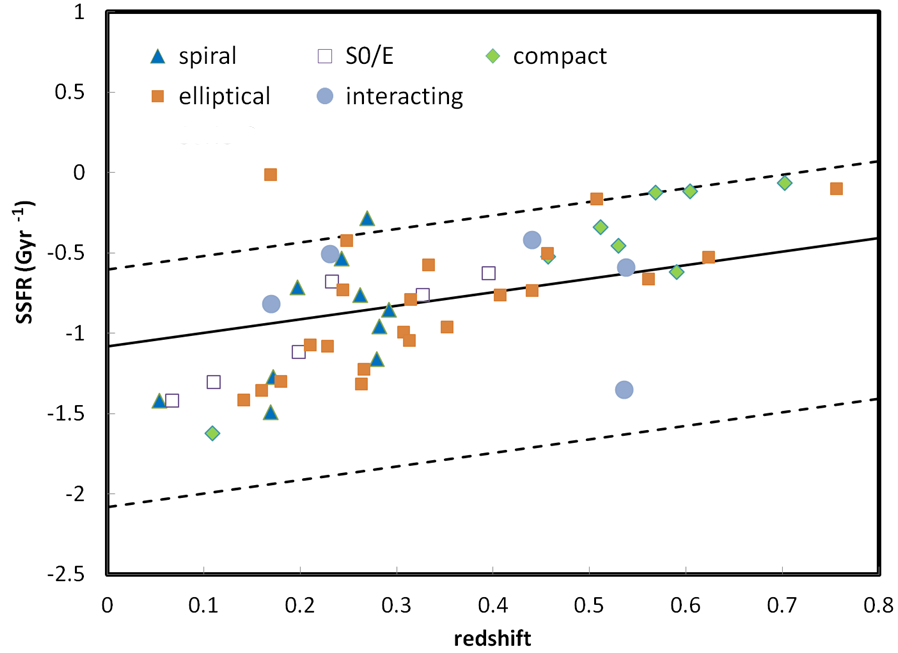}
\caption{The SSFR versus
redshift for 
\herschel-detected Type-2 AGNs with valid morphology classifications.
The solid line is the SSFR of main
sequence galaxies (Elbaz, et al. 2011).
The dotted lines are  a factor of three times or  10\% of the SSFR of main sequence galaxies, respectively.
Different symbols represent different galaxy types.
}
\label{fig:morphology}
\end{figure}

\begin{figure}[!hbt]
\centering
\includegraphics[width=0.8\textwidth]{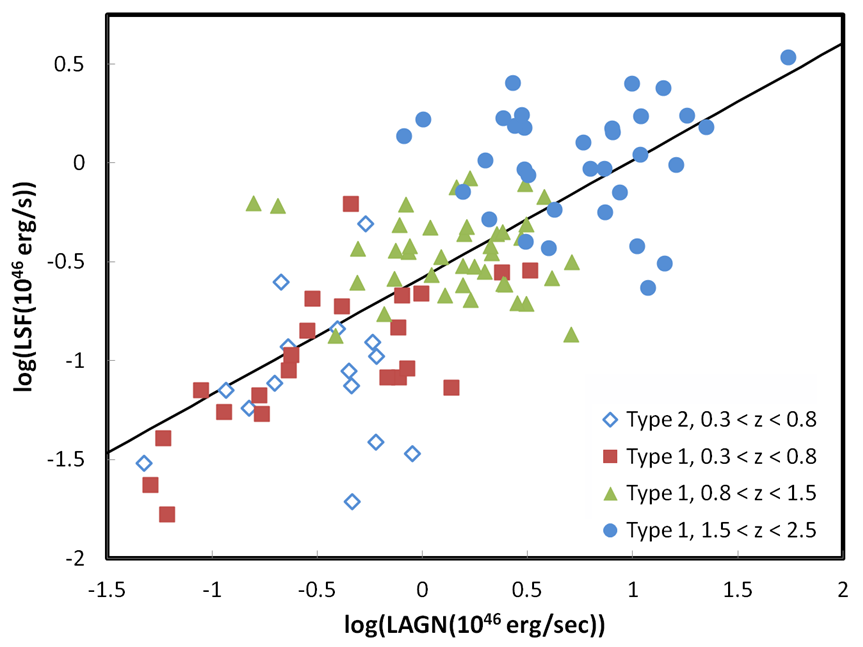}
\caption{
The relation between the IR luminosity of the star formation component
and the  AGN total luminosity.
The solid line is the best-fitting unweighted  relationship for all \herschel-detected type 1 AGN 
plus the Comparison Sample of type 2 AGN (selected to have nuclear properties
similar to those of the type 1 sources). The fit has a slope of 0.59.
}
\label{fig:agn_sf}
\end{figure}

\begin{figure}[!hbt]
\centering
\includegraphics[width=0.7\textwidth, angle =90]{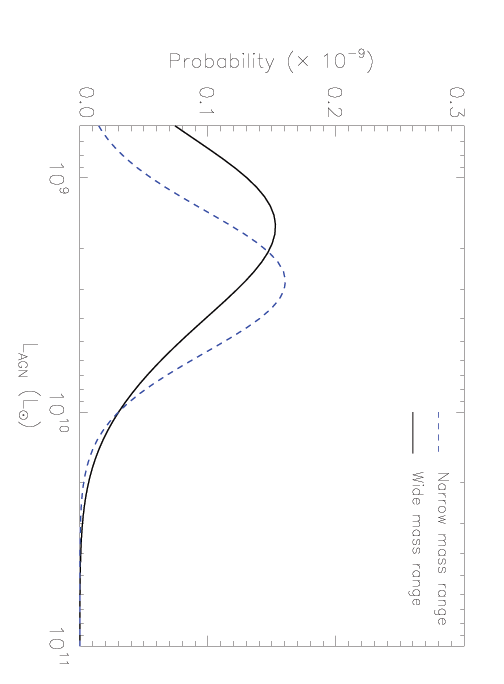}
\caption{Comparison of the AGN luminosity probability distribution 
for two different host galaxy mass distributions. 
The blue dashed line represents the probability distribution for 
the case where the stellar masses all lie within a narrow range. 
The black line represents the distribution for the case
where the stellar masses are spread over a range of a factor of ten.
For programs that detect only a small fraction of the sample 
(e.g. for $L_{\rm AGN} > 10^{10} ~\rm L_\odot$ in this example), the detection 
rate canl be much higher for the case where the masses are more widely spread.}
\label{fig:lagn_pro}
\end{figure}

\end{document}